\documentclass[amsmath, amssymb,10pt,aps,prb,twocolumn,longbibliography,notitlepage,showpacs,superscriptaddress]{revtex4-1}

\usepackage{graphicx}
\usepackage{calc}
\usepackage{braket}
\usepackage{ulem}   
\usepackage{slashed}
\usepackage{wasysym}
\usepackage{dsfont}
\usepackage{amsthm,amsmath,amsfonts,amssymb,verbatim,color}
\usepackage{subfigure}
\usepackage{bm}
\usepackage{epsfig,slashed}
\usepackage[T1]{fontenc}
\usepackage[colorlinks=true,citecolor=blue,linkcolor=blue,urlcolor=blue]{hyperref}
\usepackage{flafter}

\newcommand{\sign}{\mathrm{sign}}

\newcommand{\Ni}{N_\text{i}}
\newcommand{\kb}{k_\text{B}}
\newcommand{\kf}{k_\text{F}}

\newcommand{\vbg}{V_\text{bg}}
\newcommand{\vtg}{V_\text{tg}}
\newcommand{\dswal}{\Delta \sigma_\text{WAL}}
\newcommand{\vcnp}{V_\text{CNP}}
\newcommand{\kfb}{\overline{\kf}}
\newcommand{\rr}{\mathbf{r}}
\newcommand{\sh}{\hat{\sigma}}
\newcommand{\jj}{\mathbf{j}}
\newcommand{\del}{\mathbf{\nabla}}

\newcommand{\be}{\begin{equation}}
\newcommand{\ee}{\end{equation}}

\begin{document}
\title{Signatures of long-range-correlated disorder in the magnetotransport of ultrathin topological insulators}

\author{D. Nandi}
\thanks{These two authors contributed equally.}
\affiliation{Department of Physics, Harvard University, Cambridge, MA 02138, USA}
\affiliation{Francis Bitter Magnet Lab and Plasma Science and Fusion Centre, Massachusetts Institute of Technology, Cambridge, MA 02139, USA}
\author{B. Skinner}
\thanks{These two authors contributed equally.}
\affiliation{Department of Physics, Massachusetts Institute of Technology, Cambridge, MA 02139, USA}
\author{G.H. Lee}
\affiliation{Department of Physics, Harvard University, Cambridge, MA 02138, USA}
\affiliation{Department of Physics, Pohang University of Science and Technology, Pohang 790-784, Republic of Korea}
\author{K.-F. Huang}
\affiliation{Department of Physics, Harvard University, Cambridge, MA 02138, USA}
\author{K. Shain}
\affiliation{Department of Physics, Harvard University, Cambridge, MA 02138, USA}
\author{Cui-Zu Chang}
\affiliation{Francis Bitter Magnet Lab and Plasma Science and Fusion Centre, Massachusetts Institute of Technology, Cambridge, MA 02139, USA}
\affiliation{Department of Physics, The Pennsylvania State
University, University Park, PA 16802-6300, USA }
\author{Y. Ou}
\affiliation{Francis Bitter Magnet Lab and Plasma Science and Fusion Centre, Massachusetts Institute of Technology, Cambridge, MA 02139, USA}
\author{S.-P. Lee}
\affiliation{Department of Physics, University of Alberta, Edmonton, Alberta T6G 2E1, Canada}
\affiliation{Department of Physics and Astronomy, The John Hopkins University, Baltimore, Maryland 21218, USA }
\author{J. Ward}
\affiliation{Department of Physics, Harvard University, Cambridge, MA 02138, USA}
\author{J.S. Moodera}
\affiliation{Francis Bitter Magnet Lab and Plasma Science and Fusion Centre, Massachusetts Institute of Technology, Cambridge, MA 02139, USA}
\affiliation{Department of Physics, Massachussetts Institute of Technology, Cambridge, MA 02139, USA}
\author{P. Kim}
\affiliation{Department of Physics, Harvard University, Cambridge, MA 02138, USA}
\author{B.I. Halperin}
\affiliation{Department of Physics, Harvard University, Cambridge, MA 02138, USA}
\author{A. Yacoby}
\affiliation{Department of Physics, Harvard University, Cambridge, MA 02138, USA}
\email{yacoby@physics.harvard.edu}
\date{\today}


\begin{abstract}

In an ultrathin topological insulator (TI) film, a hybridization gap opens in the TI surface states, and the system is expected to become either a trivial insulator or a quantum spin Hall insulator when the chemical potential is within the hybridization gap.  Here we show, however, that these insulating states are destroyed by the presence of a large and long-range-correlated disorder potential, which converts the expected insulator into a metal.  We perform transport measurements in ultrathin, dual-gated topological insulator films as a function of temperature, gate voltage, and magnetic field, and we observe a metallic-like, non-quantized conductivity, which exhibits a weak antilocalization-like cusp at low magnetic field and gives way to a nonsaturating linear magnetoresistance at large field. We explain these results by considering the disordered network of electron- and hole-type puddles induced by charged impurities. We argue theoretically that such disorder can produce an insulator-to-metal transition as a function of increasing disorder strength, and we derive a condition on the band gap and the impurity concentration necessary to observe the insulating state.  We also explain the linear magnetoresistance in terms of strong spatial fluctuations of the local conductivity, using both numerical simulations and a theoretical scaling argument.

\end{abstract}

\maketitle

\section{Introduction}

Three-dimensional topological insulators (TIs) are an exotic state of matter in which gapless electronic excitations exist at the surface of a bulk system with gapped conduction and valence bands.\cite{hsieh2009tunable}  These surface states exhibit a number of interesting phenomena associated with their linear dispersion and spin-momentum locking, \cite{hasan2010colloquium} including magnetic monopole responses to applied electric field \cite{qi2009inducing} and a strong magnetoelectric effect. \cite{essin2009magnetoelectric}  Angle-resolved photoemission spectroscopy (ARPES) measurements have identified gapless Dirac surface states in several materials, including Bi$_{1-x}$Sb$_{x}$, Bi$_2$Se$_3$, Sb$_2$Te$_3$, Bi$_2$Te$_3$. \cite{hasan2010colloquium}

When a TI crystal is made very thin, however, the nature of the surface states undergoes a significant change.  In such ultrathin TI films, electrons have a finite amplitude for quantum tunneling between the top and bottom surfaces, resulting in a hybridization gap for the surface states whose magnitude depends on the film thickness $d$.  Under the right conditions, this gap can stabilize the quantum spin Hall state, which is characterized by one-dimensional, helical edge states around the border of the TI surface (as has been observed in the 2D TI HgTe/CdTe \cite{konig2007quantum,roth2009nonlocal}).

One can therefore expect a basic dichotomy of possibilities for an undoped, ultrathin TI film.  Either the system becomes a trivial insulator, with a vanishing conductivity in the limit of zero temperature, or it becomes a quantum spin Hall insulator, with a quantized conductance.  The fate of the TI film vis-a-vis these two possibilities is predicted to depend in a nontrivial way on the value of the thickness $d$, with the system oscillating between a quantum spin Hall and a trivial insulating state as a function of thickness.\cite{liu2010oscillatory}

In this paper, however, we find evidence for a third possibility, outside of this dichotomy, in which a hybridization gap exists but the insulating state is destroyed by the presence of long-range-correlated disorder.  We measure the resistivity of ultrathin films of pristine (Bi,Sb)$_2$Te$_3$ as a function of temperature, chemical potential, and magnetic field, and we find a number of features that suggest that a dominant role is played by long-range-correlated disorder, which arises inevitably due to charged impurities in the film and in the substrate.  Our films are 4 quintuple layers (QLs) thick, which is predicted to produce a quantum spin Hall insulator, but we find instead a finite, nonquantized value of the resistance in the limit of zero temperature.  We provide an explanation for this observation in terms of an insulator-to-metal transition produced by increasing long-ranged disorder.  We also observe a prominent weak anti-localization correction and a large linear magnetoresistance at high magnetic fields, which we explain in terms of strong spatial fluctuations of the local conductivity.

The remainder of this paper is organized as follows.  In Sec.\ \ref{sec:sample} we briefly describe our sample preparation and measurement setup.  Section \ref{sec:B0} describes our zero-field measurements, and presents theoretical arguments to explain a resistivity that is both finite and non-quantized in the limit of zero temperature.  Section \ref{sec:MR} presents results for the resistivity as a function of magnetic field and gate voltage, along with a theoretical discussion of both the weak antilocalization corrections and the linear magnetoresistance that we observe.

\section{Sample preparation and measurement setup}
\label{sec:sample}

Our TI films are made from 4 QLs of (Bi,Sb)$_2$Te$_3$ grown on a SrTiO$_3$(111) substrate using molecular-beam epitaxy in an ultra-high vacuum. Each QL layer is $\sim 1$\,nm thick. The Bi, Sb and Te effusion cells as well as the SrTiO$_3$ (111) substrate are held at high temperature in order to ensure precise control of surface stoichiometry. The crystallinity of the films is monitored by reflection high-energy electron diffraction (RHEED) pattern. To prevent oxidation in ambient conditions, $2$\,nm of Tellurium and $2$\,nm of Alumina capping layer is grown on top of the films. The samples tend to degrade at high temperatures, and hence all the processing was done at no higher than 100 $^\circ$C. Contacts were made using e-beam lithography. In the dual gated devices, the top gate dielectric was made from 20 nm of HfO$_2$ grown by atomic layer deposition.\cite{Nandi2018} Magneto-transport measurements are performed in a dilution refrigerator with a 8 T magnet and using low frequency ac lockin techniques. Results in this paper are taken from five Hall bar devices, which we denote H$_1$, H$_2$, H$_3$, H$_4$ and H$_5$  .

\section{Transport at zero magnetic field}
\label{sec:B0}

\begin{figure}[htb!]
\begin{center}
\includegraphics[width=0.7\columnwidth]{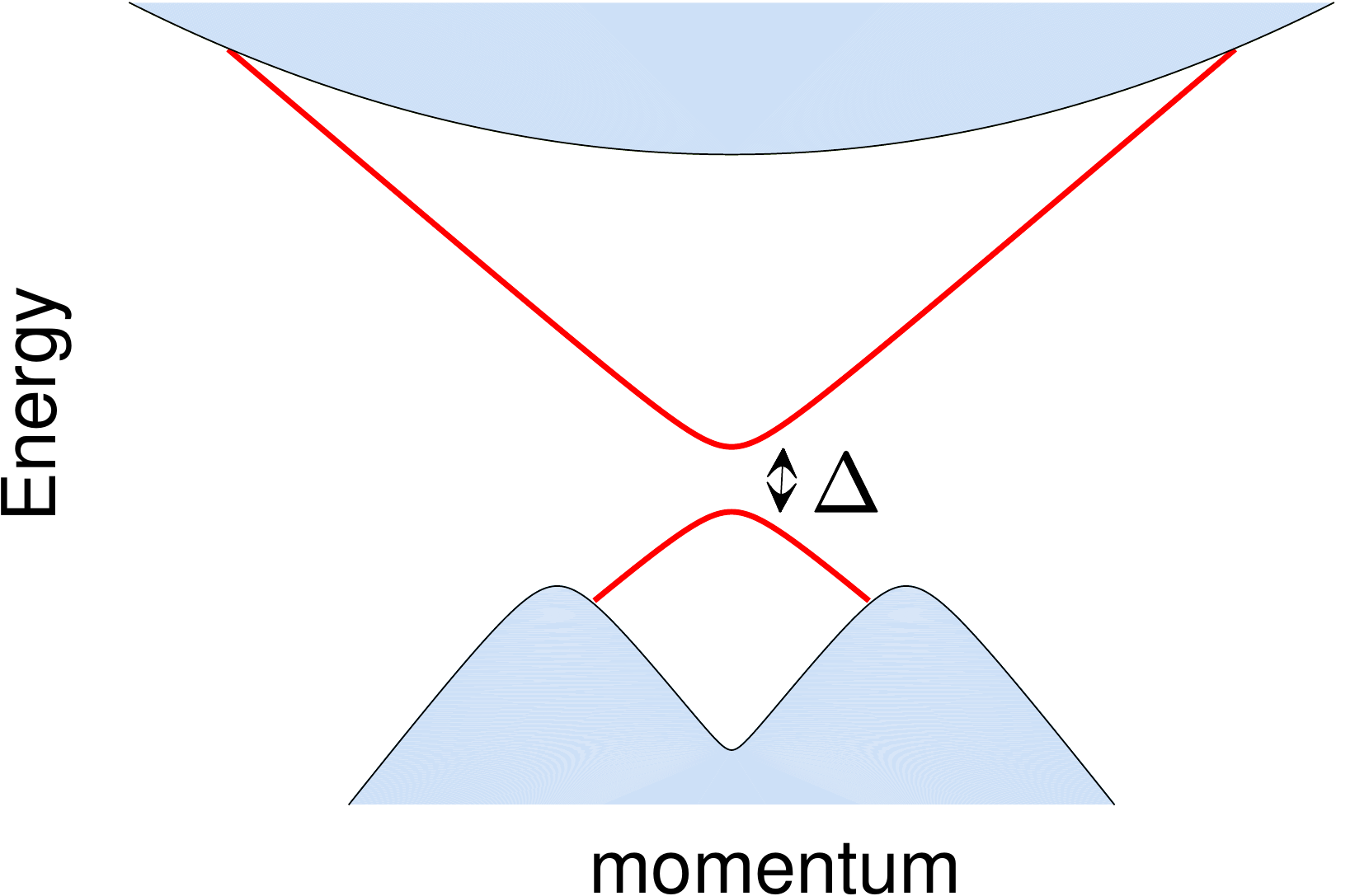}
\caption{Schematic picture of the dispersion of the Dirac surface states (red lines).  Hybridization between top and bottom surfaces opens a small gap $\Delta$ at the $\Gamma$ point.  The bulk conduction and valence band states are denoted by the upper and lower blue shaded areas, respectively.}
\label{fig:bands}
\end{center}
\end{figure}

As mentioned in the Introduction, hybridization between the two parallel TI surfaces leads to a gap opening at the Dirac point of the surface dispersion relation (as illustrated in Fig.\ \ref{fig:bands}).  The size $\Delta$ of this gap is generally expected to be between 5\,meV and 50\,meV; this range encompasses estimates from density functional theory and tight-binding models for the gaps in 4-QL-thick Bi$_2$Se$_3$ and Bi$_2$Te$_3$.\cite{liu2010oscillatory} For situations where the chemical potential resides in the middle of the gap and there is no band bending, one would expect the TI surface to become an insulator with an activation energy $\Delta/2$ for the conductivity.  Such a state would have either zero conductivity in the limit of zero temperature (if the system is a trivial insulator) or a quantized conductance $2 e^2/h$ (if the system forms a quantum spin Hall state).

\begin{figure}[htb!]
\begin{center}
\includegraphics[width=\columnwidth]{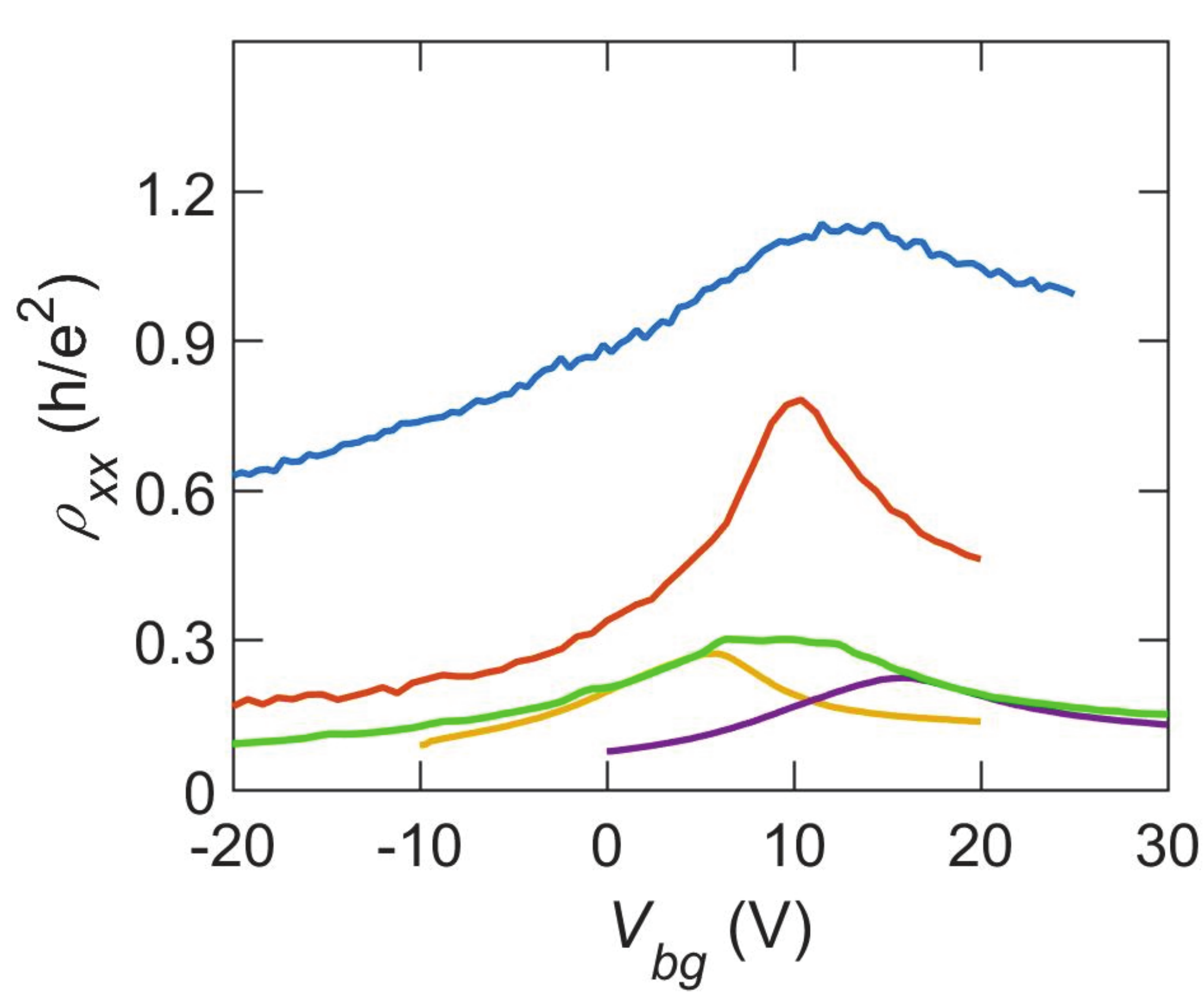}
\caption{Sheet resistivity of 5 different Hall bar devices --- denoted H$_1$ (blue curve),  H$_2$ (brown), H$_3$ (green), H$_4$ (yellow) and  H$_5$ (purple) --- as a function of back gate voltage.   The top gate voltage is held fixed at $V_{tg}$ = 0. The measurement  temperature was $\sim$ 30 mK. }
\label{fig:minimal_resistivity}
\end{center}
\end{figure}

Using a dual-gated FET setup, we shift the chemical potential for both the top and bottom surfaces of our samples across a wide range in order to search for these insulating states. At large negative voltages the chemical potential resides far below the energy of the Dirac point, while at large positive voltages the chemical potential is high above.  When the two gate voltages are chosen such that both surfaces are at the charge-neutral point (CNP), the system assumes its maximally insulating state.  This behavior is shown in Fig.\ \ref{fig:minimal_resistivity} for five different samples as a function of the back gate voltage. (The behavior as a function of both back and top gate voltages is discussed in the Supplementary Material.)

\begin{figure}[htb!]
\begin{center}
\includegraphics[width=\columnwidth]{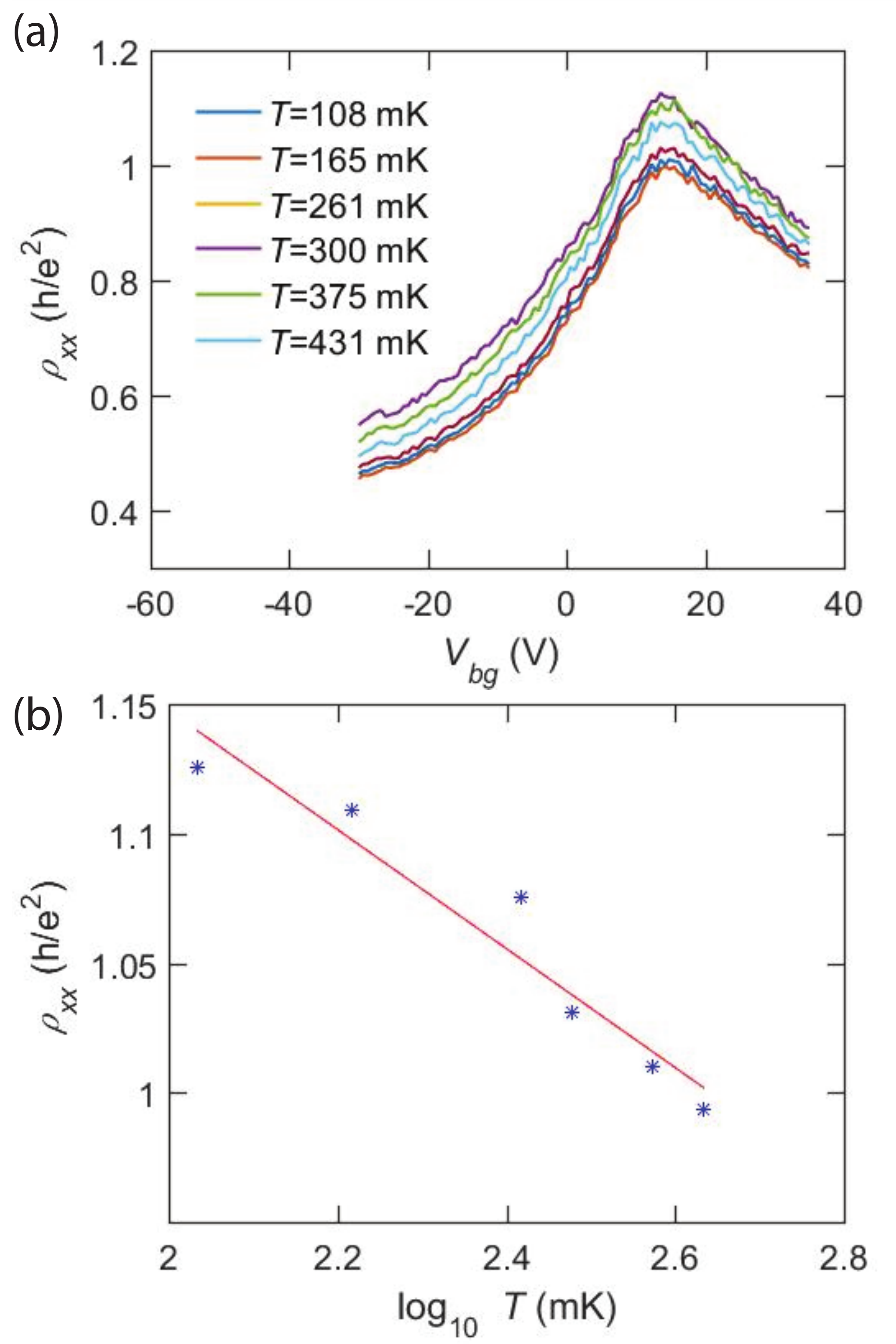}
\caption{(a) The longitudinal resistivity $\rho_{xx}$ of device  H$_1$ as a function of back gate voltage for different values of the temperature. (b) The maximal resistivity (corresponding to the CNP) of device H$_1$ is plotted as a function of temperature, showing only a weak, logarithmic dependence on temperature.
 }
\label{fig:Tdependence}
\end{center}
\end{figure}

Contrary to the expectation for a clean system, our measurements reveal a conductivity that is neither insulating-like nor quantized.  Indeed, Fig.\ \ref{fig:minimal_resistivity} shows that the resistance takes a value of order $h/e^2$ at the CNP, but this value varies from one sample to the next.  As shown in Fig.\ \ref{fig:Tdependence}, the resistivity depends very weakly on temperature, even though the temperature is far below $\Delta/\kb \gtrsim 50$\,K. Using the traditional description of an undoped semiconductor, one would predict an activated  dependence $\rho_{xx} \propto \exp[\Delta/(2 \kb T)]$.

Taken together, these two observations suggest that the system is not well described by either the clean band insulator or quantum spin Hall insulator states. The most trivial explanation for our results would be that the surface bands simply do not have a band gap.  For example, in 4-QL-thick films of Bi$_2$Te$_3$ the $\Gamma$ point lies at a local minimum of the valence band, \cite{yazyev2010spin} and consequently there is no finite window of energy with zero density of states. In Sb$_2$Te$_3$, tunneling experiments have shown that films of the same thickness have a thermodynamic gap that is no larger than a few meV. \cite{jiang2012landau}  Still, it is worth considering whether there is another, more interesting explanation for the lack of insulating behavior, especially since similar results have been recorded for transport in thin films of Bi$_2$Se$_3$, \cite{kim2011thickness} despite a Dirac point that lies well outside the bulk valence band states and a gap $\Delta$ on the order of tens of meV. \cite{liu2010oscillatory, zhang2010crossover}

\begin{figure}[htb!]
\begin{center}
\includegraphics[width=0.9\columnwidth]{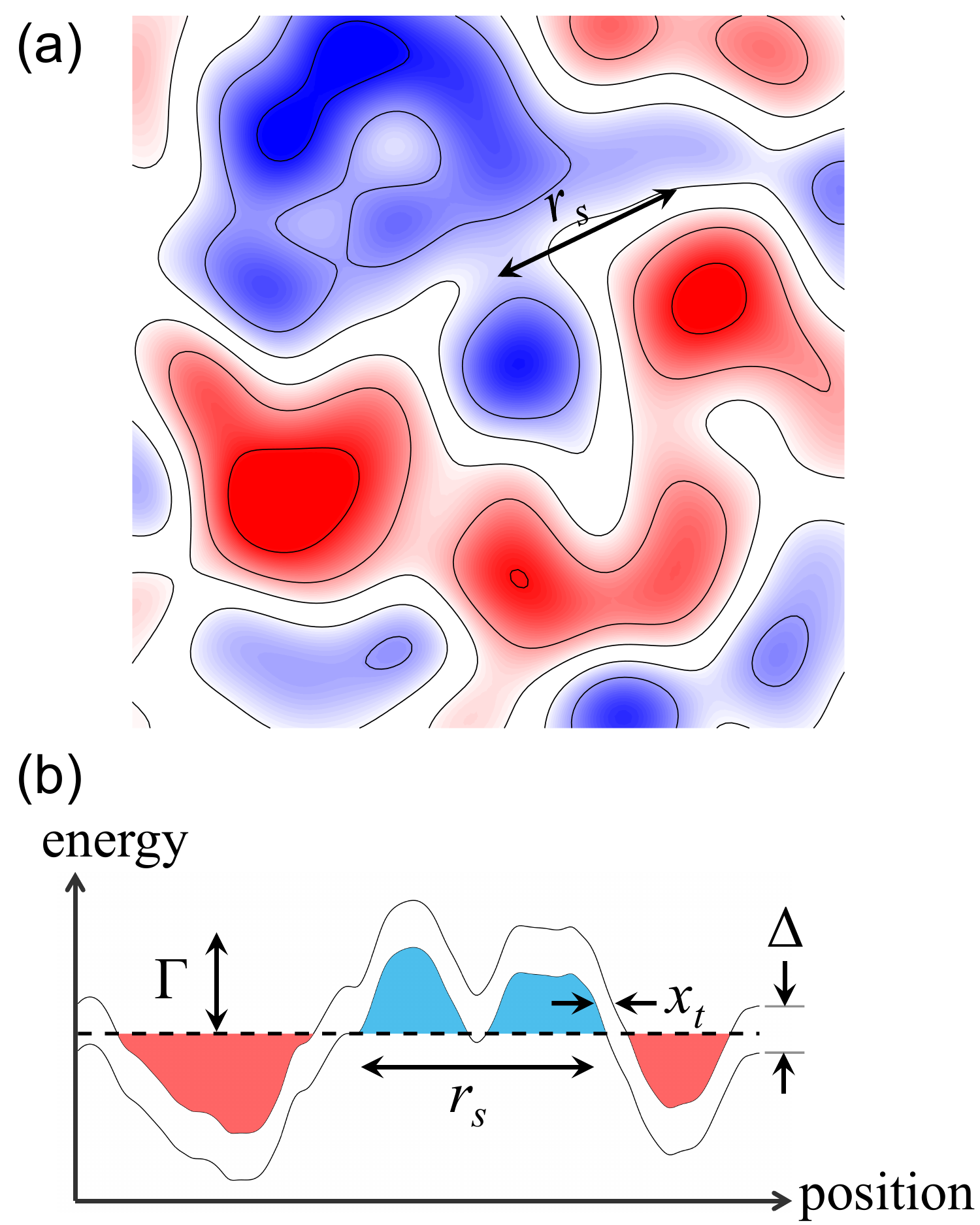}
\caption{(a) Schematic illustration of the disorder potential landscape.  Charged impurities produce a slowly-varying Coulomb potential that locally creates puddles of electrons (red) and holes (blue).  The typical correlation length of the potential, $r_s$, is labeled, and thin black lines show contours of constant potential.  White regions denote tunnel barriers between electron and hole puddles, which exist at large enough band gap $\Delta$.
(b) A schematic illustration of band bending, showing the energy along some particular direction on the surface.  The surface band gap $\Delta$ is labeled, along with the typical magnitude $\Gamma$ of the disorder potential and the width $x_t$ of the tunnel barrier between electron and hole puddles.}
\label{fig:slowpotential}
\end{center}
\end{figure}

The apparent breakdown of the clean insulator picture can be rationalized by considering the effects of long-ranged disorder induced by Coulomb impurities, which exist both in the TI film and in the substrate.  Such impurities are known to provide long-wavelength fluctuations of the local Fermi energy, which provide a finite density of states at zero energy due to band-bending. \cite{skinner2013effects}  This random band-bending effect is illustrated in Fig.\ \ref{fig:slowpotential}.  When the surface gap $\Delta$ is sufficiently large, the fluctuations of the Fermi energy lead to the formation of isolated electron and hole puddles [red and blue regions of Fig.\ \ref{fig:slowpotential}(a), respectively], separated by insulating tunnel barriers (white regions).  When the chemical potential $\mu$ is precisely in the middle of the band gap (which we define as $\mu = 0$), electron and hole puddles appear in equal number. Similar picture of fluctuating Fermi energy in the presence of insulating gap has been also adopted in graphene nanoribbons \cite{stampfer2009energy} and dual-gated bilayer graphene \cite{lee2015continuous}. One can estimate the condition for maintaining a good insulating state at $\mu = 0$ by demanding that the typical tunneling action $S$ for electron tunneling across such a barrier satisfy $S \gg \hbar$.

To estimate the action $S$, we first assume that the typical magnitude $\Gamma$ of the disorder potential is sufficiently large that $\Gamma \gg \Delta$.  We also assume that the typical correlation length $r_s$ of the potential is much larger than the film thickness.  Both of these assumptions are validated below.
Under these assumptions one can use numerical estimates for $\Gamma$ and $r_s$ based on a gapless TI surface. \cite{skinner2013theory}$^{,}$ \footnote{The theoretical estimates of Ref.\ \onlinecite{skinner2013theory} are slightly modified for our case, since the two opposite TI surfaces screen the disorder in tandem and therefore the density of states should be doubled relative to the case of a single surface.}  These estimates give
\be
\Gamma \simeq \left(\frac{\pi^3 \alpha_s^2}{2} \right)^{1/6}
\hbar v \Ni^{1/3}
\label{eq:Gamma}
\ee
and
\be
r_s \simeq \frac{\Ni^{-1/3}}{(2 \alpha_s)^{4/3}},
\label{eq:rs}
\ee
where $\Ni$ is the (three-dimensional) concentration of impurities in the TI film and substrate, $\hbar$ is the reduced Planck constant, $v$ is the Dirac velocity, and $\alpha_s = e^2/(4\pi\varepsilon_0 \varepsilon \hbar v)$ is the effective fine structure constant, with $\varepsilon$ the effective dielectric constant and $\varepsilon_0$ the vacuum permittivity. If one assumes an impurity concentration of order $10^{19}$\,cm$^{-3}$, a Dirac velocity $v$ of order $5 \times 10^{5}$\,m/s, and an effective dielectric constant $\varepsilon$ as large as several hundred (due to the close proximity of the highly-polarizable SrTiO$_3$ substrate), then $\alpha_s$ is of order $0.1$, the disorder potential $\Gamma$ is of order $20$\,meV, and $r_s \sim 40$\,nm.  It is worth noting that at chemical potentials far from the CNP, both $r_s$ and $\Gamma$ will generally be smaller than their $\mu = 0$ values.

In order to estimate the width $x_t$ of the typical spatial separation between electron and hole puddles, one can notice that the typical in-plane electric field is $F \sim \Gamma/(e r_s)$, so that $x_t$ is given by $e F x_t \sim \Delta$.  Solving for $x_t$ and substituting Eqs.\ (\ref{eq:Gamma}) and (\ref{eq:rs}) gives $x_t \sim 4 \pi \varepsilon_0 \varepsilon \Delta/(\alpha_s^{2/3} e^2 \Ni^{2/3})$.  The typical tunneling action between electron and hole puddles can be estimated as the product of the height $\sim \Delta$ of the tunnel barrier and the time $\sim x_t/v$ needed to traverse it.  So $S \sim \Delta x_t/v \sim \hbar \Delta^2/[\alpha_s^{5/3} (\hbar v \Ni^{1/3})^2]$.  For the system to be insulating, one must have $S \gg \hbar$, which is equivalent to
\be
\Delta \gg \alpha_s^{5/6} \hbar v \Ni^{1/3}.
\label{eq:Deltareq}
\ee
Equation (\ref{eq:Deltareq}) can be viewed as a generic requirement for the existence of an insulating state in a gapped 2D system on a substrate with charged impurities.  That is, either the gap $\Delta$ must be large enough or the impurity concentration $\Ni$ must be small enough that Eq.\ (\ref{eq:Deltareq}) is satisfied.

In our samples, the right-hand side of Eq.\ (\ref{eq:Deltareq}) is of order $10$\,meV.  For much smaller values of the gap, one can say that electron and hole puddles are well-connected by quantum tunneling, and there is no meaningful ``insulating barrier'' between them.  Our samples apparently correspond to such a situation, where Eq.\ (\ref{eq:Deltareq}) is violated, so that one can think of the surface as effectively ungapped even though $\Delta$ is finite.  Producing a well-insulating TI thin film apparently requires either a larger hybridization gap $\Delta$ or a much smaller impurity concentration $\Ni$.  For the remainder of this paper we set $\Delta = 0$ when discussing transport.

The zero-field DC resistivity is given by $\rho = (h/e^2) /(\kf \ell)$, where $\kf$ is the typical Fermi wave vector and $\ell$ is the electron mean free path.  For the ``puddled'' scenario depicted in Fig.\ \ref{fig:slowpotential}, the typical value of $\kf$ at zero chemical potential is $\sim  \alpha_s^{1/3} \Ni^{1/3}$, while the mean free path is of the same order as $r_s$.  Thus, the resistivity at zero chemical potential is given by \cite{skinner2013theory}
\be
\rho_{\text{max}} \simeq  \frac{h}{e^2} \alpha_s \ln(1/\alpha_s).
\ee
For our samples, this expression gives a value of order $\approx 0.3 h/e^2$.

Our picture of conduction through a spatially-disordered landscape is also consistent with measurements of the superconducting proximity effect, which we present in the Supplementary Material.

\section{Magnetotransport}
\label{sec:MR}

\begin{figure}[htb!]
\begin{center}
\includegraphics[width=\columnwidth]{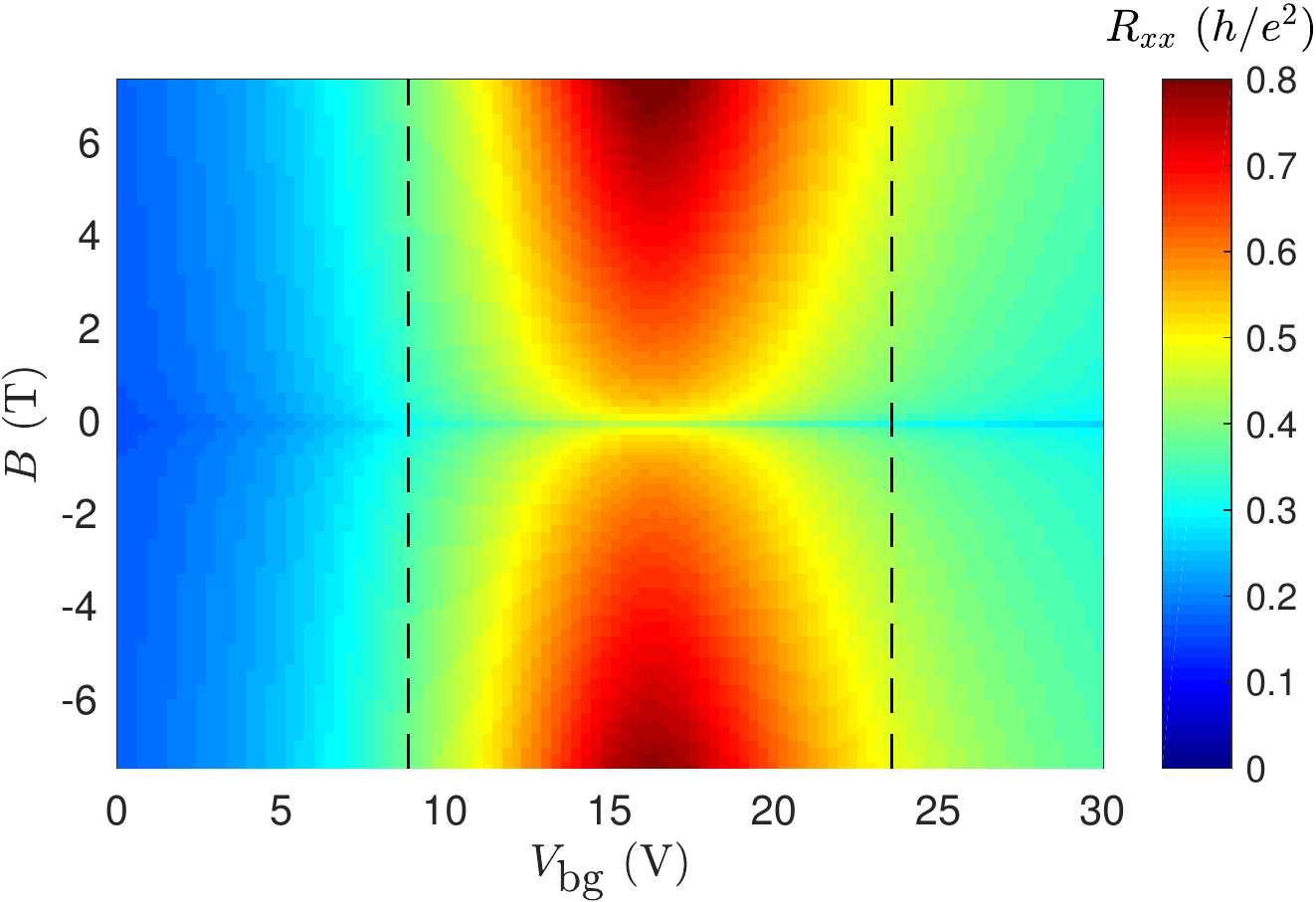}
\caption{Longitudinal resistance $R_{xx}$ of device H$_5$ as a function of back gate voltage $\vbg$ and magnetic field $B$.  The measurement temperature $T= 50$\,mK. $R_{xy}$ is everywhere much smaller than $R_{xx}$. Vertical dashed lines demarcate the three regimes of voltage depicted in Fig.\ \ref{fig:WAL}(a)--(c).}
\label{fig:MR}
\end{center}
\end{figure}

We also study the electron transport under the application of a perpendicular magnetic field.  The measured longitudinal resistance $R_{xx}$ is plotted in Fig.\ \ref{fig:MR} for device H$_4$ as a function of the field strength $B$ and the back gate voltage $\vbg$.  For a given magnetic field, the resistance is maximized near the CNP, which for this device corresponds to $\vbg \approx 16$\,V.  The resistance also increases monotonically as a function of $B$. In general, we observe an asymmetry between positive and negative voltages relative to the CNP, with negative voltages generally corresponding to smaller resistance.  This asymmetry suggests that negative values of the chemical potential correspond to a larger density of states than positive values of the chemical potential, which may arise either because of curvature of the Dirac band or because of proximity of the Dirac point to the bulk valence band states (as depicted in Fig.\ \ref{fig:bands}).  In our measurement conditions, the Hall resistance $R_{xy}$ is everywhere much smaller than the longitudinal resistance $R_{xx}$; this is shown explicitly in the Supplementary Material.  Thus we can approximate the conductivity $\sigma \simeq (L/w)/R_{xx}$, where $L/w \approx 2$ is the aspect ratio of the sample H$_5$.  The leading order correction to this expression is of order $(R_{xy}/R_{xx})^2$, which is smaller than $5\%$ throughout the regime of our measurements.

Our data show two notable features as a function of magnetic field. For any given gate voltage there is a sharp cusp in $R_{xx}(B)$ near $B = 0$, which previous experimental studies have attributed to weak antilocalization (WAL).\cite{steinberg2011electrically, liu2012crossover, tian2014quantum, zhang2013weak, kim2011thickness} The correction $\dswal$ to the conductivity associated with WAL is described by the theory of Hikami, Larkin, and Nagaoka: \cite{hikami1980spin}
\be
\dswal = \frac{\alpha}{\pi} \frac{e^2}{h} \left[ \psi\left( \frac{1}{2} + \frac{\hbar}{4 e B L_\phi^2} \right) - \ln \left( \frac{\hbar}{4 e B L_\phi^2} \right) \right].
\label{eq:WAL}
\ee
Here $\psi(z)$ is the digamma function, $L_\phi$ is the phase coherence length, and $\alpha$ is a numerical coefficient defined so that $\alpha < 0$ indicates WAL and $\alpha > 0$ corresponds to weak localization.  The quantity $2|\alpha|$ is usually associated with the number of parallel conduction channels.

\begin{figure}[htb!]
\begin{center}
\includegraphics[width=\columnwidth]{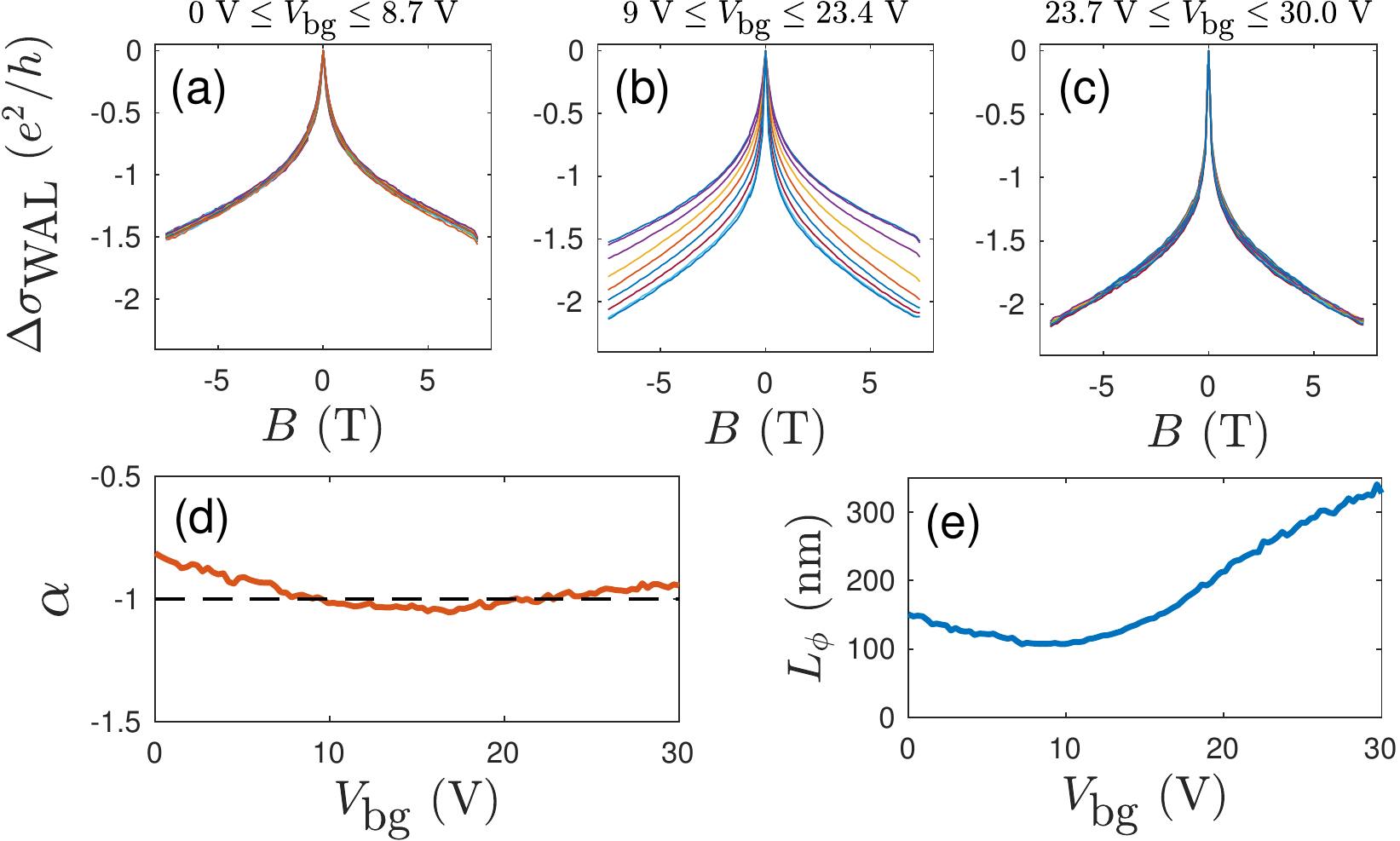}
\caption{(a)--(c) The measured WAL correction to the conductivity of device H$_5$ as a function of magnetic field for different ranges of the back gate voltage $\vbg$ (the range of $\vbg$ is indicated in the title of each plot, and is equivalent to the three ranges demarcated in Fig.\ \ref{fig:MR}).  For each plot the different curves correspond to different values of $\vbg$.  (d) The value of the constant $\alpha$ extracted from a fit to Eq.\ (\ref{eq:WAL}) as a function of $\vbg$.  (e) The extracted phase coherence  length as a function of $\vbg$.  The measurement temperature is $T = 50$\,mK. }
\label{fig:WAL}
\end{center}
\end{figure}

In Fig.\ \ref{fig:WAL} we plot $\dswal = \sigma(B) - \sigma(0)$ as a function of magnetic field for different values of the back gate voltage.  The sharp, logarithmic cusp of $\dswal(B)$ is consistent with Eq.\ (\ref{eq:WAL}), and we can perform good fits in the range $-2\,\textrm{T} < B < 2\,\textrm{T}$ in order to extract the parameters $\alpha$ and $L_\phi$.  Interestingly, for low enough voltages that the chemical potential is far below the CNP, we find that all measured curves $\dswal(B)$ are identical, irrespective of the gate voltage [Fig.\ \ref{fig:WAL}(a)].  Similarly, all measured values of $\dswal(B)$ for voltages far above the CNP also collapse onto a single curve [Fig.\ \ref{fig:WAL}(c)], although this curve is distinct from the one corresponding to low voltages.  In the intermediate voltage range $9 \textrm{ V} \leq \vbg \leq 23.4 \textrm{ V}$ (between the two dashed vertical lines in Fig.\ \ref{fig:WAL}), the behavior of $\dswal(B)$ transitions smoothly from one limiting curve to the other [Fig.\ \ref{fig:WAL}(b)].  We interpret these two limiting curves as corresponding to chemical potentials either far below or far above the Dirac point.  In the latter case, the current is carried only by the linear Dirac surface states above the gap.  In the opposite limit of small $\vbg$, the current moves through an admixture of Dirac surface states and conduction band states.  For intermediate voltages, the disorder potential mixes these two behaviors spatially by random band bending.

By fitting our data to Eq.\ (\ref{eq:WAL}), we are able to extract estimates for the constant $\alpha$ and the phase coherence length $L_\phi$ as a function of the back gate voltage.  These results are shown in Fig.\ \ref{fig:WAL}(d) and (e), respectively.  It is worth remarking that the inferred value of $\alpha$ is everywhere close to $-1$, as one might expect for a conduction process with two parallel channels (arising from the two parallel surfaces).  The estimated phase coherence length is on the order of several hundred nm, consistent with previous studies at low temperature.\cite{steinberg2011electrically, tian2014quantum}

\begin{figure}[htb!]
\begin{center}
\includegraphics[width=\columnwidth]{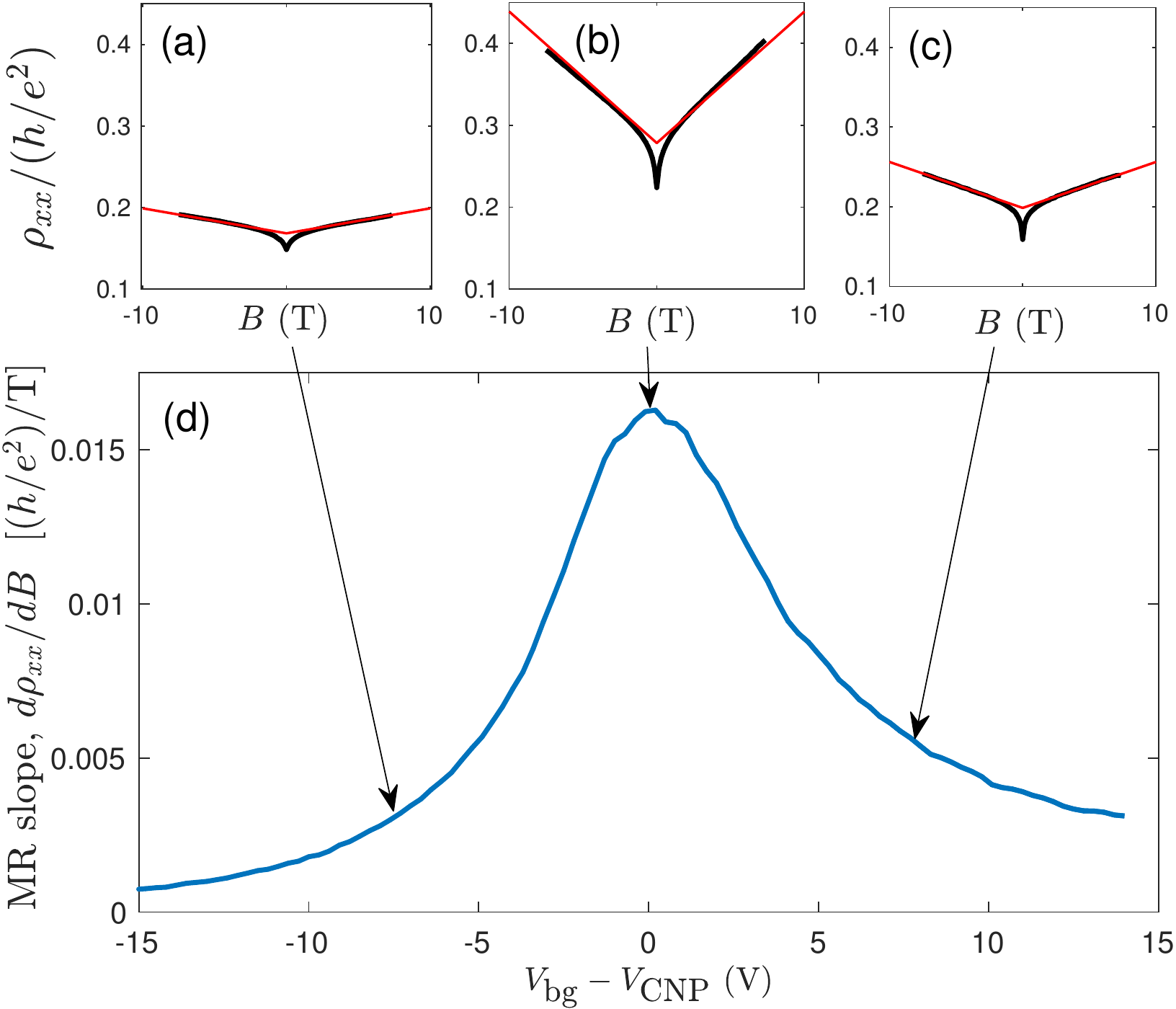}
\caption{ (a) -- (c) The resistance $R_{xx}$ of Hall bar device H$_5$ as a function of magnetic field is shown for different values of the back gate voltage, whose value for each plot is indicated by the arrow pointing downward.  The thick black line in each plot shows the experimental data, and the thin red line is a straight-line guide to the eye that indicates the linear slope of $R_{xx}$ versus $|B|$.
(d) The value of the slope $dR_{xx}/dB$ of device H$_5$ at large $B$ is plotted as a function of back gate voltage $\vbg$.  $\vcnp \approx 16$\,V indicates the voltage at the charge neutral point.}
\label{fig:B0}
\end{center}
\end{figure}

At larger magnetic field, the WAL correction gives way to a resistance that increases linearly with magnetic field strength, with no evidence of saturation.  As shown in Fig.\ \ref{fig:B0}, this linear magnetoresistance (LMR) effect is most prominent near the CNP.  As the chemical potential is moved away from the CNP in either direction, the slope of the LMR is reduced.

Multiple explanations have been proposed during the last few decades for nonsaturating LMR in 2D electron systems.
For example, Wang and Lei have proposed a mechanism for LMR on a TI surface based on Zeeman splitting.\cite{wang2012linear}  However, the magnitude of the LMR associated with this mechanism is much smaller than the value we observe.  Indeed, to explain our largest observed LMR slope with the Zeeman splitting mechanism would apparently require an electron $g$-factor of several hundred, which seems inconsistent with transport experiments in a tilted magnetic field. \cite{chen2011tunable}
Other authors have explored more generic, semiclassical explanations for LMR, and have shown how it can arise from either persistent gradients of electron density \cite{ilan2006longitudinal} or mesoscopic spatial fluctuations of the mobility. \cite{parish2005classical, kozlova2012linear}  Such fluctuations are commonly treated using either resistor network models \cite{parish2005classical} or effective medium approaches \cite{Knap2014transport, Ping2014disorder}, which give largely equivalent results. \cite{Ramakrishnan2017equivalence}

Similar to these latter approaches, we suggest a way to understand our results based on a simple model in which the system is described by a local, Drude-like conductivity tensor that varies as a function of position due to spatial fluctuations in the electron density. In particular, we suppose that one can define local longitudinal and Hall conductivities $\sigma_{xx}(\rr)$ and $\sigma_{xy}(\rr)$, respectively, which vary as a function of position $\rr$.  Such a description is generally valid so long as variations in the electron density occur over a length scale $r_s$ that is much longer than the mean free path $\ell$ or the Fermi wavelength $\sim \kf^{-1}$.

In the Drude model, the ratio $\sigma_{xy}/\sigma_{xx} = \omega_c \tau$, where $\omega_c$ is the cyclotron frequency and $\tau$ is the transport scattering time.  For a gapless Dirac system
\be
\omega_c \tau = \frac{e B \ell}{\hbar \kf},
\ee
where $\kf$ is the local value of the Fermi momentum and $\ell = v \tau$ is the local mean free path.  (As declared above, we are again ignoring the effects of any small band gap $\Delta$.)  Note that the ratio $\sigma_{xy}/\sigma_{xx}$ becomes large when $B$ is sufficiently large.  The value of the local Fermi momentum can be described by the Thomas-Fermi equation,
\be
E_\text{F}[\kf(\rr)] - e \phi(\rr) = \mu,
\label{eq:TF}
\ee
where $E_\text{F} = \hbar v \kf \times \sign(\mu + e \phi)$ is the Fermi energy relative to the Dirac point and $\phi(\rr)$ is the electrostatic potential.

In order to understand the appearance of LMR within this model, consider first the case when the system is close enough to the CNP that electron and hole puddles exist in almost equal number, $|\mu| \ll \Gamma$.  This is the regime where the LMR is observed to be most prominent experimentally, and one can understand its appearance using the following scaling arguments.  Near the CNP, electron and hole puddles are nearly equally abundant, and the local conductivity at the boundary between puddles is small because of the locally vanishing value of the electron density.  Consequently, the current across the system is forced to pass through narrow ``pinch points'' of the random potential, where the electric potential is close to zero and adjacent electron- or hole-type puddles are narrowly separated.  These pinch points provide the bottleneck for conduction, and they become more prominent with increasing magnetic field. \cite{Ruzin1993Hall} (See the Supplementary Material for simulated images of current flow.) One can think that an order unity number of such pinch points exist per square area $\sim r_s^2$, and consequently, if $G$ is the typical conductance of the pinch point, then the longitudinal resistivity of the system is $\rho_{xx} \sim 1/G$.

To estimate the typical conductance $G$ of the pinch point, one can exploit the result for the (two-terminal) conductance of a square with nonzero Hall conductivity\cite{hans1958, hojgaard1972geometrical, abanin2008conformal}: $G_\square = \sqrt{\sigma_{xx}^2 + \sigma_{xy}^2}$.  If the magnetic field $B$ is large enough, then $\sigma_{xy} \gg \sigma_{xx}$ at the pinch point and we arrive at the relation $\rho_{xx} \sim 1/|\sigma_{xy}^{(0)}|$, where $\sigma_{xy}^{(0)}$ represents the Hall conductivity at the pinch point.  At $\omega_c \tau \gg 1$ the Hall conductivity $\sigma_{xy} \simeq e n/B = \pm e \kf^2/(2 \pi B)$, and so our result for the longitudinal resistivity is equivalent to
\be
\rho_{xx} \sim \frac{h}{e^2} \frac{e B}{\hbar [\kf^{(0)}]^2},
\label{eq:MRslope}
\ee
where $\kf^{(0)}$ represents the typical Fermi momentum at the pinch point.

In general, pinch points are locations where the random potential is close to zero.  Thus, if the chemical potential is not too close to zero, then Eq.\ (\ref{eq:TF}) implies $\kf^{(0)} = |\mu|/(\hbar v)$.  As the chemical potential is shifted away from the CNP, the corresponding value of $\kf^{(0)}$ increases and the slope of the magnetoresistance declines.  This is consistent with the experimental result in Fig.\ \ref{fig:B0}.  On the other hand, as one approaches the CNP very closely, both the typical spatial size of the pinch point and its typical Fermi momentum are reduced, and the resistance increases.  While the relation $\kf^{(0)} = |\mu|/(\hbar v)$ implies a divergence of the resistance at $\mu \rightarrow 0$, such a divergence may be truncated by the finite mean free path.  In other words, since the local conductivity is not well-defined at scales shorter than the mean free path $\ell$, one can think that the minimal size of the pinch point is $\sim \ell$, and consequently that the minimal value of $\kf^{(0)}$ is $\sim \kfb \ell/r_s$, where $\kfb$ is the typical value of $\kf$ near the center of an electron or hole puddle.  Using the estimates for $\kfb$ and $\ell$ presented below, and inserting the expression for $\kf^{(0)}$ into Eq.\ (\ref{eq:MRslope}), gives a maximum magnetoresistance slope of order $\rho_{xx} \sim 0.03 (h/e^2)$ per Tesla of field.  This is consistent in order of magnitude with our measured result.

\begin{figure}[htb!]
\centering
\includegraphics[width=\columnwidth]{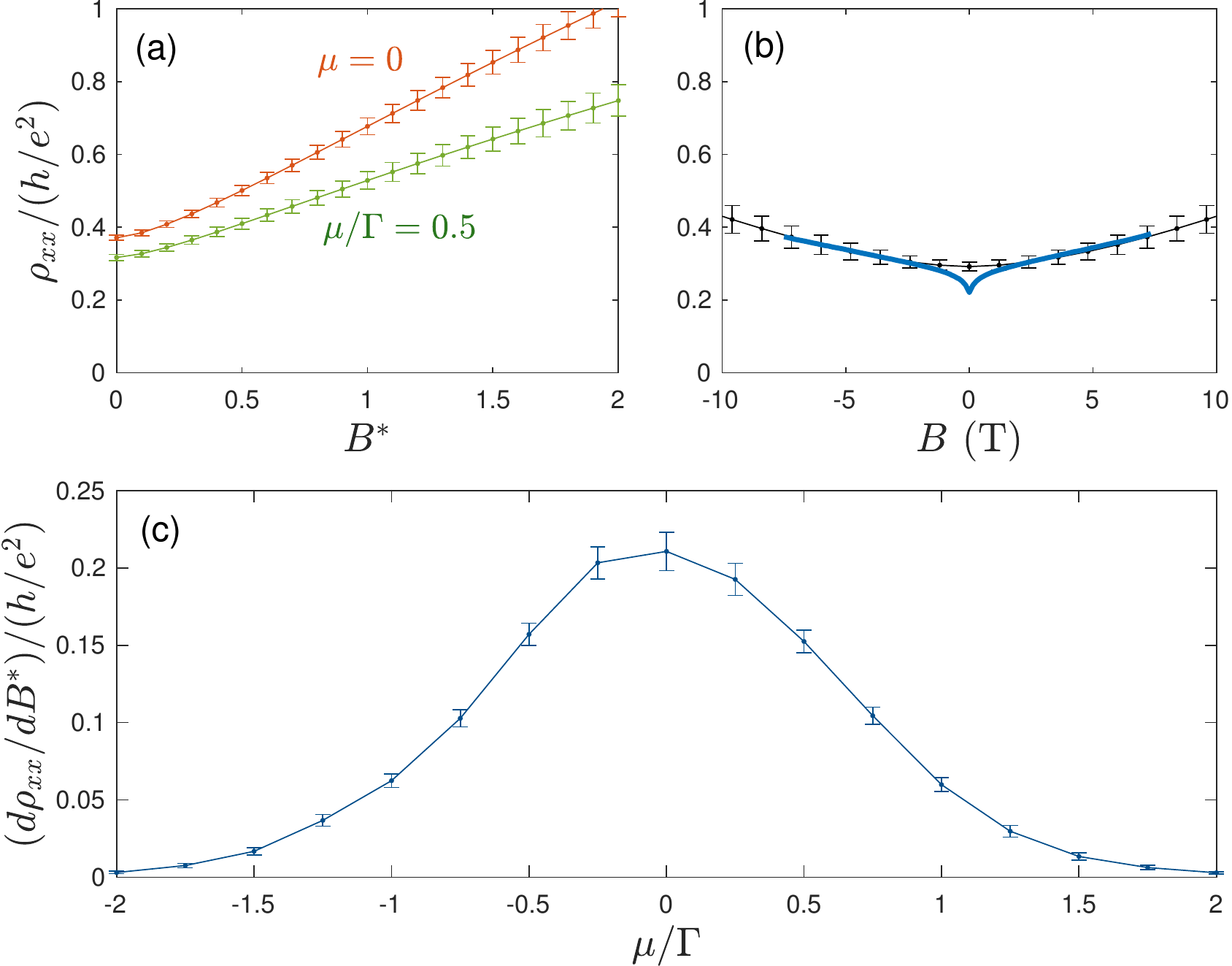}
\caption{Magnetoresistance for a simulated Hall bar with random disorder.  (a) Two example curves are plotted for $\rho_{xx}$ as a function of the dimensionless magnetic field $B^* = B \ell^2/(\hbar e)$.  Both curves correspond to $\kfb \ell = 6$, where $\kfb$ is the root-mean-square deviation of the Fermi momentum from its mean value and $\ell$ is the mean free path, which is taken to be a constant.  The curves are labeled by their corresponding value of the chemical potential $\mu$, normalized to the root-mean-square amplitude $\Gamma$ of the disorder potential.  (b) Shows a fit to the experimental data (thick blue curve, corresponding to device H$_5$) for $\rho_{xx}(B)$ at the charge-neutral point ($\mu = 0$).  The simulation data (black line with error bars) corresponds to $\kfb = 0.88$\,nm$^{-1}$ and $\kfb \ell = 6.5$. (c) The magnetoresistance slope $d\rho_{xx}/dB^*$ is plotted as a function of chemical potential $\mu$, normalized to the disorder potential amplitude $\Gamma$.  The slope is calculated by a linear fit to simulation data in the interval $0.2 < B^* < 1$, and in this example $\kfb \ell = 6$ is held constant.}
\label{fig:rhoB-sim}
\end{figure}

In order to test our scaling arguments quantitatively, we implemented numeric, finite-element simulations of current flow through a Hall bar geometry with a correlated disorder potential and a Drude-like conductivity tensor having a local value of $\kf(\rr)$ given by  Eq.\ (\ref{eq:TF}).  For simplicity, our simulations assume a transport scattering time $\tau$ that is independent of energy or position.  While there is no reason \textit{a priori} to expect this assumption to be accurate quantitatively, the scaling argument leading to Eq.\ (\ref{eq:MRslope}) suggests that the longitudinal resistivity $\rho_{xx}$ becomes independent of $\tau$ at sufficiently large field.  As shown in Fig.\ \ref{fig:rhoB-sim}, the simulation consistently reproduces the LMR trend, as well as the decline in the LMR slope with increasing chemical potential.  Details of the simulation method are provided in the Supplementary Material (along with results for $\rho_{xy}$, which are consistent with experiment).  Within the assumption of a constant scattering time, one can fit the experimental data at the CNP quantitatively by setting the root-mean-square Fermi momentum $\kfb = 0.88$\,nm$^{-1}$, and the mean free path $\ell = 7.4$\,nm.  More details about the fitting are provided in the Supplementary Material.

\section{Conclusion}

In this paper we have presented experimental results for the resistivity of thin TI films as a function of temperature, chemical potential, and magnetic field.  In the absence of disorder, these systems are predicted to form a quantum spin Hall state.  We find, however, that the transport in our system is dominated by long-range fluctuations of the disorder potential, presumably induced by charged impurities in the film and in the substrate.  In particular, such long-ranged disorder creates a random landscape of $p$- and $n$-type regions, and this landscape destroys the insulating state.  Our theoretical arguments suggest that one may reach an insulating state only when the gap is large enough and the impurity concentration is low enough that Eq.\ (\ref{eq:Deltareq}) is satisfied.

The magnetotransport shows signs of both weak antilocalization and a nonsaturating linear magnetoresistance.  The WAL correction $\dswal(B)$ is described well by the usual Hikami-Larkin-Nagaoka theory, Eq.\ (\ref{eq:WAL}), with two parallel conduction channels ($\alpha \approx -1$). We also find that $\dswal(B)$ collapses onto one of two curves when the chemical potential is far from the CNP.  We have shown that the linear magnetoresistance can be interpreted as the result of spatial fluctuations in the local conductivity arising from strong disorder fluctuations.  Our estimate of the linear MR slope and its dependence on chemical potential are both consistent with observations.

Taken together, our results provide new understanding of electron transport in ultrathin topological insulators, and may bring us closer to realizing ideal quantum spin Hall insulators.  More broadly, our improved understanding of the disorder-induced insulator-to-metal transition and LMR may be important for a wide class of disordered 2D electron systems.

\

\acknowledgments

It is a pleasure to thank I. Sodemann, A.~Nahum, S.~Simon, J.~T.~Chalker, A.~Kamenev, and Y.~P.~Chen for valuable discussions. We are indebted to Di Wei, Lucas Orona, Tony Zhou, Charlotte B{\o}ttcher, Michael Kosowsky and Andrew Pierce for invaluable help with fabrication and measurements. AY, DN, KS and JW acknowledge the support from Gordon and Betty Moore Foundation Grant No. 4531, NSF grant No. DMR-1708688, ARO Grant No.~W911NF16-1-0491, ARO Grant No.~W911NF-17-1-0023, and ARO grant No.~W911NF-18-1-0316. Fabrication for this work was supported by DOE Award Number DE-SC0001819.  BS was supported as part of the MIT Center for Excitonics, an Energy Frontier Research Center funded by the U.S. Department of Energy, Office of Science, Basic Energy Sciences under Award No.~DE-SC0001088. BIH acknowledges support from the STC Center for Integrated Quantum Materials under NSF grant DMR-1231319. PK, GHL and KH acknowledge support from NSF Grant No. DMR-1420634. JSM, CZC and YO acknowledge the support from NSF Grant No. DMR-1700137, ONR Grant No. N00014-16-1-2657 and the Center for Integrated Quantum Materials under NSF Grant No. DMR-1231319. CZC also thanks the support from Alfred P. Sloan Research Fellowship and ARO Young Investigator Program Award (W911NF1810198).

\section{Supplemental Material for `Signatures of long-range-correlated disorder in the magnetotransport of ultrathin topological insulators'}

\subsection{Hall measurements}

\begin{figure}[htb]
\begin{center}
\includegraphics[width=3.0in]{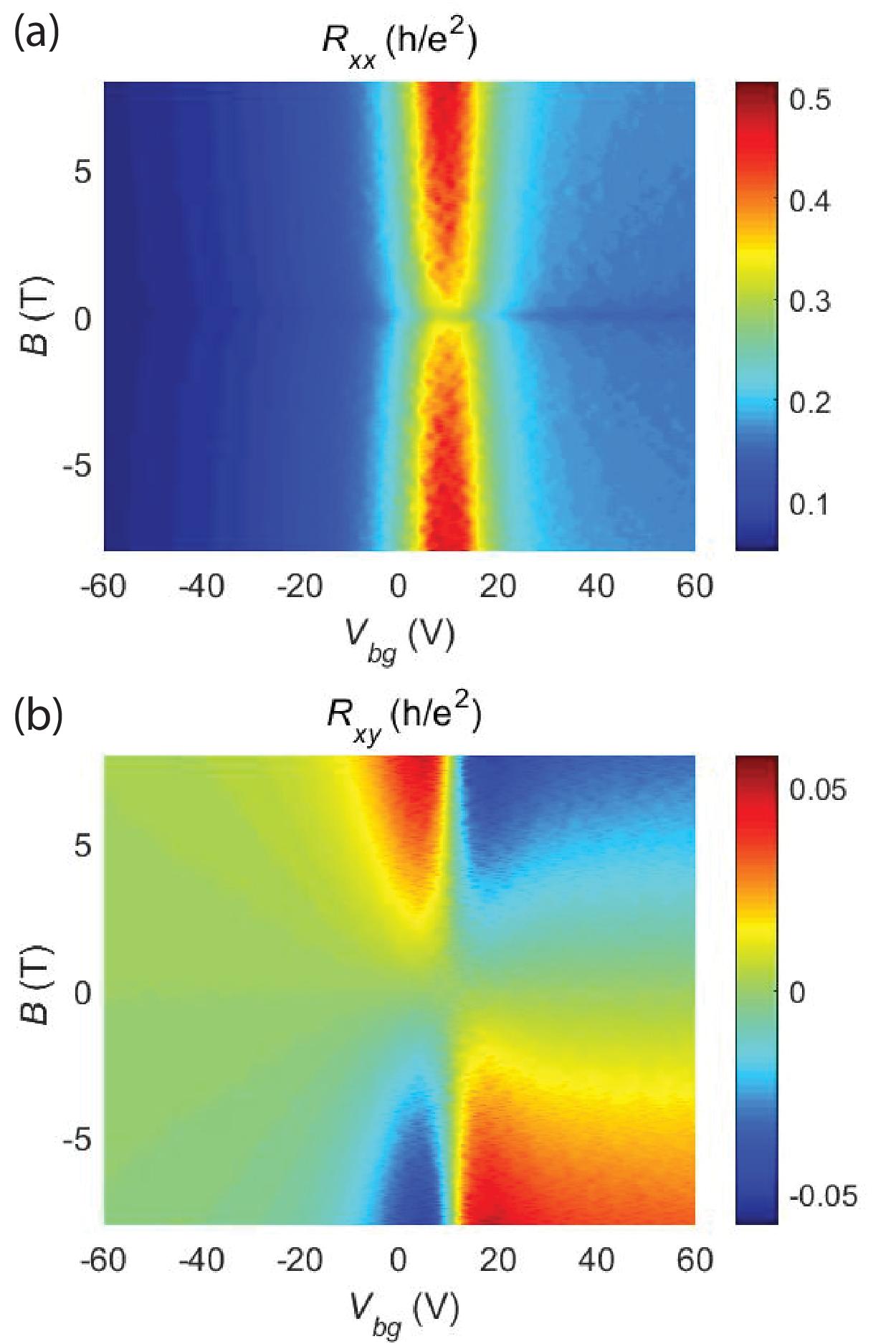}
\caption{(a) Longitudinal resistance $R_{xx}$ and (b) Transverse Hall resistance $R_{xy}$ for Hall bar device H$_3$.}
\label{Rxy}
\end{center}
\end{figure}

In the main text we focused on the longitudinal resistance $R_{xx}$ and ignored the transverse Hall resistance $R_{xy}$.  Here we present results for $R_{xy}$, and we show that it is everywhere much smaller than $R_{xx}$.  In Fig.\ \ref{Rxy} we plot $R_{xx}$ and $R_{xy}$ for device H$_3$ as a function of back gate voltage $\vbg$ and magnetic field strength $B$.

\begin{figure}[htb]
\begin{center}
\includegraphics[width=3.0in]{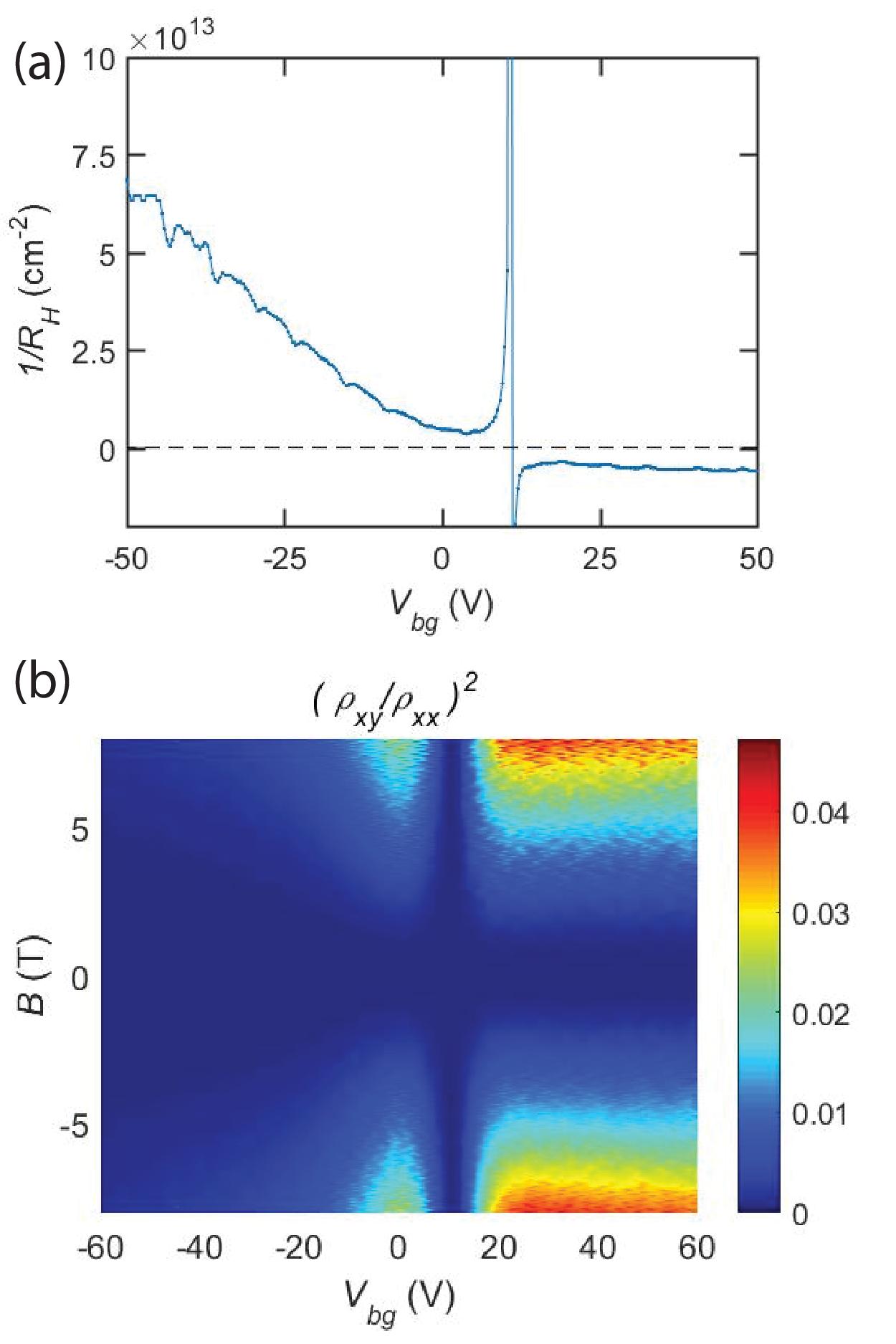}
\caption{(a) Inverse Hall coefficient as a function of back gate voltage. (b) Ratio of $(\rho_{xy}/\rho_{xx})^2$ plotted as a function of magnetic field and backgate voltage. Data corresponds to device H$_3$, which has an aspect ratio $L/w \approx 1$.}
\label{Ratio}
\end{center}
\end{figure}

From the $R_{xy}$ data, we calculate the inverse Hall coefficient $1/R_H$, as shown in Fig.\ \ref{Ratio}(a). The strong asymmetry in $R_H$ on different sides of the charge neutral point (CNP) reflects an asymmetry in the band structure of the TI films.  The minimum value of $|1/R_H| \sim 3 \times 10^{12}$\,cm$^{-2}$ suggests a typical carrier density of electron and hole puddles.  Similar results for $R_H$ were seen in Ref.\ \onlinecite{lee2015mapping}.

In the presence of a perpendicular magnetic field, the conductivity is given in general by
\begin{align}
\sigma_{xx} & =\frac{\rho_{xx}}{(\rho_{xx}^2+\rho_{xy}^2)} \nonumber \\
& = \frac{1}{\rho_{xx}(1+ \rho_{xy}^2/\rho_{xx}^2)}.
\end{align}
Figure \ref{Ratio}(b) shows that for all relevant magnetic field and backgate voltages,  the quantity $(\rho_{xy}/\rho_{xx})^2 \ll 1$. Hence, the global conductivity can be well approximated by $\sigma_{xx}\approx 1/\rho_{xx}$.

\subsection{Dual Gating}

\begin{figure}[htb]
\begin{center}
\includegraphics[width=3.0in]{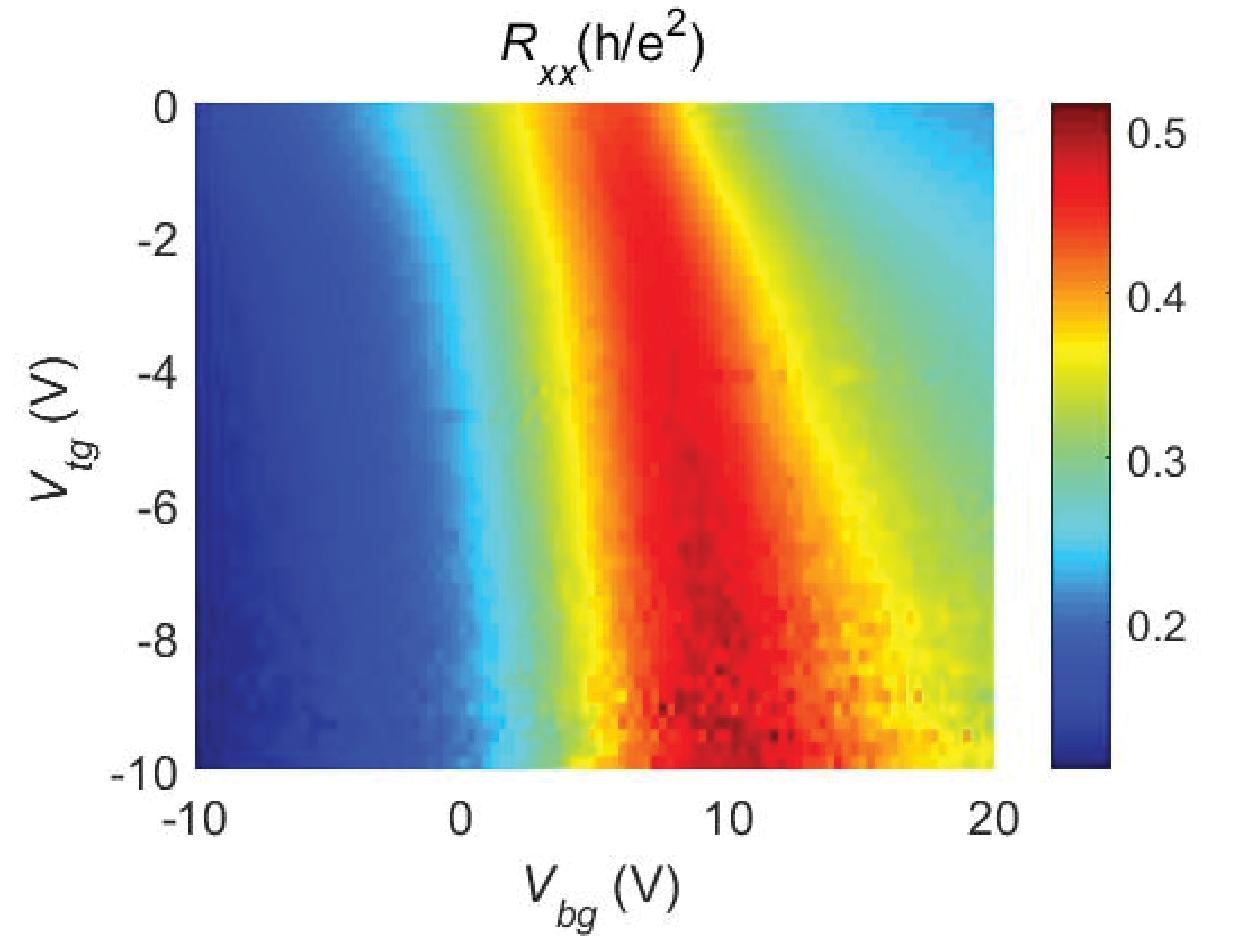}
\caption{Dependence of the longitudinal resistance $R_{xx}$ on top and backgate voltages for device H$_4$.}
\label{Dual Gating}
\end{center}
\end{figure}

In some of our samples, both a back gate and a top gate were fabricated, allowing the chemical potential to be modulated by two independent gate voltages.  In Fig.\ \ref{Dual Gating} we present the low-temperature longitudinal resistance $R_{xx}$ for sample H$_4$ as a function of the top and back gate voltages, $\vtg$ and $\vbg$, respectively. For this sample, the longitudinal resistance exhibits a maximum of $\approx 0.5 h/e^2$ at the CNP, and there is no sign of an insulating state with large resistance.

\subsection{Superconducting proximity effect}

\begin{figure}[h!]
\begin{center}
\includegraphics[width=3.0in]{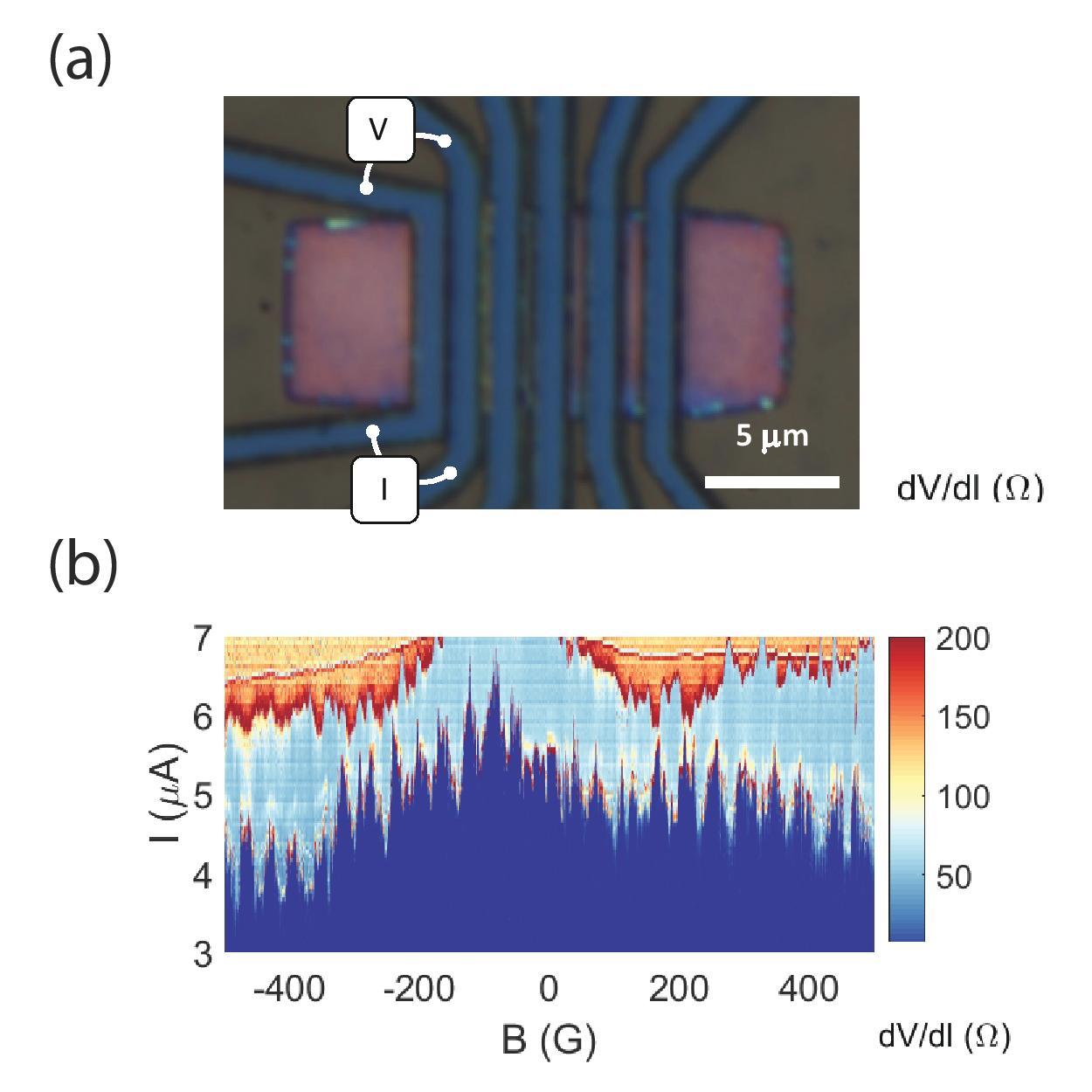}
\caption{(a) Optical image and  (b) Fraunhoffer pattern of device JJ1.}
\label{Josephson junction}
\end{center}
\end{figure}

Coupling superconductivity to topological insulator (TI) surface states is predicted to create a p$_x$+ip$_y$ superconductor with Majorana bound states.\cite{fu2008superconducting,fu2008superconducting,tanaka2009manipulation,potter2013anomalous} Such Majorana states may enable topologically protected qubits for quantum computation,\cite{moore2010birth} and consequently there has been an intensive search for Majorana modes in exfoliated 3D TIs \cite{williams2012unconventional,rokhinson2012fractional,kurter2015evidence}. Achieving Josephson coupling is an important step towards engineering topological superconductivity.

We  investigated superconducting Josephson coupling mediated by the TI thin film, as depicted schematically in Fig.\ \ref{Josephson junction}(a). An important consideration for achieving Josephson coupling is a transparent superconducting contact with minimum contact resistance. To make transparent contacts, we etched Tellurium capping layer covering topological insulator film (Bi,Sb)$_2$Te$_3$, and in-situ evaporated Ti (5 nm) / Nb (2.5 nm)/ NbN (50 nm) without breaking vacuum in the evaporation chamber. The device dimensions were 6 $\mu$m $\times$ 0.1 $\mu$m, as shown in Fig.\ \ref{Josephson junction}(b).

The interference pattern of the critical current $I_c$($B$) of the Josephson junction under perpendicular magnetic field $B$ gives valuable information about the Josephson current density $J_c$($x$) along the junction width. In the case of a spatially uniform Josephson current density, the interference pattern corresponds to a single slit Fraunhofer pattern where the lobes of $I_c$ decays with $1/B$. On the other hand, for TIs in which conduction occurs dominantly at the edges of the system, the interference pattern is expected to resemble the double slit-like diffraction pattern of a dc SQUID with non-decaying lobes. In this case the critical current $I_c (B)$ is given by
\be
I_c (B)=I_c \left| \cos\left(\frac{\pi A B} {\phi_0} \right) \right|,
\ee
where $\phi_0$ is the magnetic flux quantum and $A$ is the area of the junction. Similar analyses of the interference pattern of Josephson junctions has shown edge dominant conduction in HgTe/HgCdTe 2D topological insulator \cite{hart2014induced} and Bi$_{1.5}$Sb$_{0.5}$Te$_{1.7}$Se$_{1.3}$ 3D TIs. \cite{lee2014local}
\begin{figure*}[htb!]
\centering
\includegraphics[width=16cm]{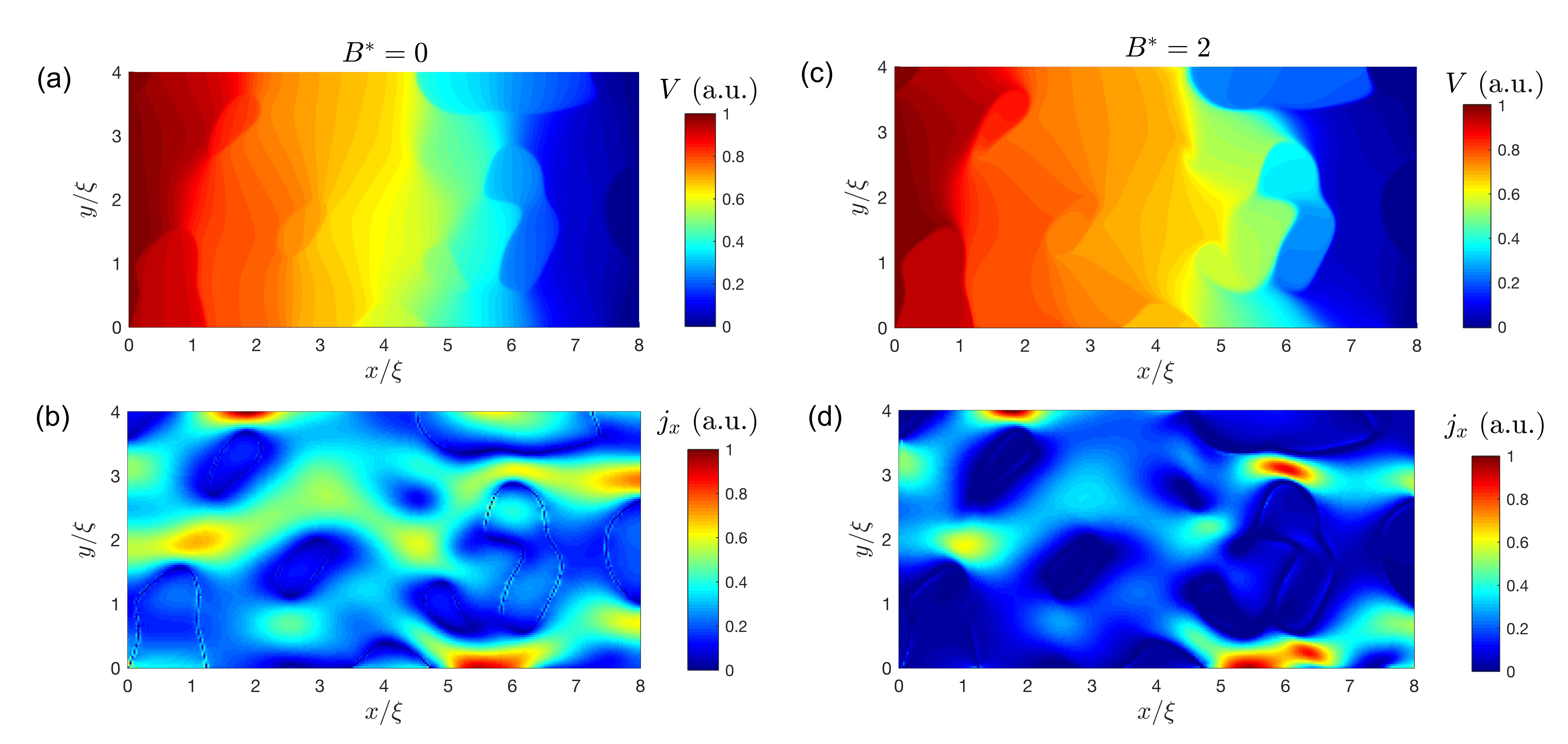}
\caption{Example numeric solutions for the electrochemical potential $V$ and the current $j_x$ in the $x$-direction for a given realization of the random potential $P(x,y)$. (a) and (b) correspond to zero magnetic field, and (c) and (d) correspond to the same random potential with a field $B^* = 2$. (b) and (d) have the same color scale.  All images correspond to zero chemical potential, $\mu = 0$.}
\label{fig:current_examples}
\end{figure*}
In our devices, the normal junction resistance $R_N \approx 60$\,$\Omega$ and the maximum Josephson current $I_c \approx 7 \mu$A at the temperature of 50 mK give $I_c R_N \approx 420 \mu$eV, which corresponds to about $\Delta_\text{sc}/6e$, where $\Delta_\text{sc}$ is the superconducting gap. The Josephson junction displays an interference pattern, as shown in Fig.\ \ref{Josephson junction}(c), that cannot be explained by a uniform Josephson current distribution. The extracted period of the quasi-periodic oscillation of $I_c$ is about 35 G, which matches well with the expected value of $\phi_0/A = 34$\,G.
 Imperfect constructive and destructive interferences visible in the pattern are indicative of the effect of random disorder in the junction. Proximity-induced Josephson coupling and the similar interference pattern were also observed by us in Josephson junctions made of Ti (5 nm)/Al (50 nm).

\begin{figure}[h]
\centering
\includegraphics[width=8cm]{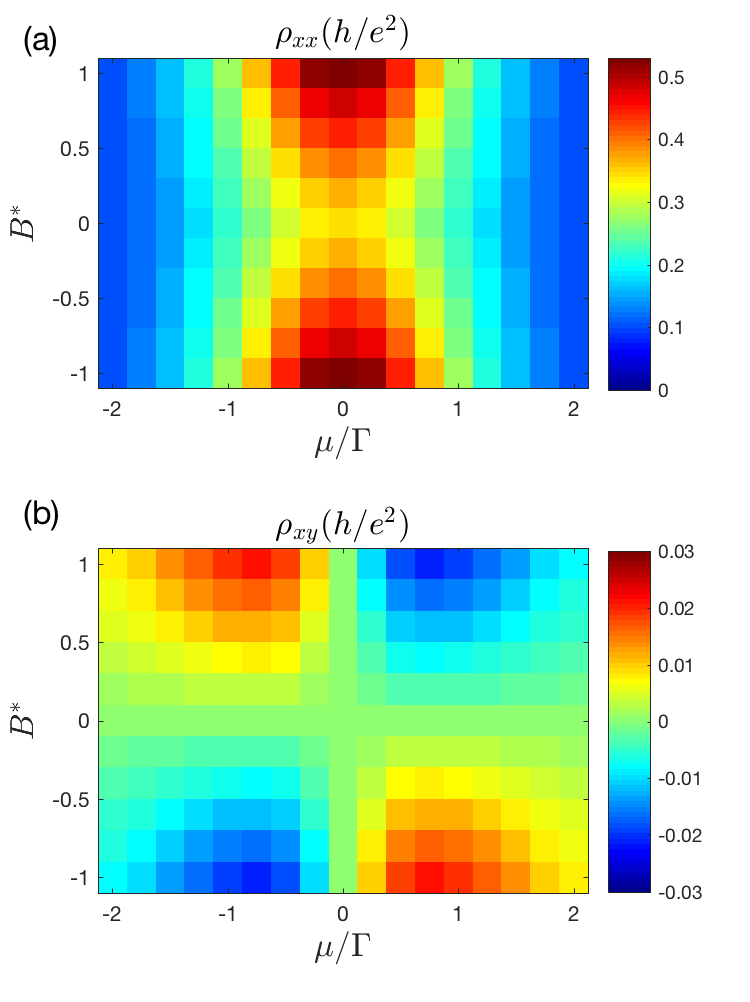}
\caption{Simulated values of the longitudinal (a) and Hall (b) resistivities as a function of magnetic field and chemical potential. The vertical axis corresponds to the dimensionless magnetic field $B^*$, while the horizontal axis corresponds to the chemical potential $\mu$ in units of the root-mean-square disorder potential $\Gamma$.  In this example $\kfb \ell = 6$ everywhere.  Compare to Fig.\ \ref{Rxy}, and note that the scale of $\rho_{xy}$ is much smaller than $\rho_{xx}$.}
\label{fig:rhoxx-xy}
\end{figure}
\subsection{Numeric simulations of current flow}

In the main text we presented results for the magnetoresistance based on a model where the local conductivity tensor $\sh(\rr)$ varies with the position $\rr$ due to spatial variations in the electron/hole concentration.  Here we present more details about our numeric simulations of the resistivity.

Within the Drude model, the conductivity tensor has the form
\be
\sh(\rr) = \left( \begin{array}{cc}
\sigma_{xx}(\rr) & -\sigma_{xy}(\rr) \\
\sigma_{xy}(\rr) & \sigma_{xx}(\rr) \end{array} \right),
\ee
where the values of $\sigma_{xx}(\rr)$ and $\sigma_{xy}(\rr)$ are related to the value of the local electrostatic potential by
\begin{eqnarray}
\sigma_{xx}(\rr) & = & \frac{e^2}{h} \kf(\rr) \ell \frac{1}{1 + [\omega_c(\rr) \tau]^2}, \nonumber \\
\sigma_{xy}(\rr) & = & \frac{e^2}{h} \kf(\rr) \ell \frac{\omega_c(\rr) \tau}{1 + [\omega_c(\rr) \tau]^2},
\label{eqs:sigma}
\end{eqnarray}
where $\kf(\rr)$ is the local value of the Fermi momentum, $\omega_c(\rr)$ is the local value of the cyclotron frequency, $\tau$ is the transport scattering time and $\ell = v \tau$ is the mean free path.  In principle, both $\ell$ and $\tau$ may have a dependency on the local Fermi energy, and therefore on position.  However, for simplicity in our numeric simulations we take $\ell$ and $\tau$ to be constants with no spatial variation.  The local Fermi momentum $\kf(\rr)$ is related to the disorder potential $\phi(\rr)$ by the Thomas-Fermi equation, Eq.\ (7) of the main text.

In order to address the problem computationally, we define the following dimensionless units.  First, we define a normalized electrochemical potential
\be
P(\rr) = \frac{e \phi(\rr) + \mu}{\sqrt{ \langle ( e \phi)^2 \rangle_\rr } },
\ee
so that $P(\rr)$ has a standard deviation of unity.  Here, $\mu$ is the chemical potential and $\sqrt{ \langle ( e \phi)^2 \rangle_\rr } \equiv \Gamma$ is the root-mean-square value of the disorder potential.  The local value of the Fermi momentum is then given by
\be
\kf (\rr) = \kfb \left| P(\rr) \right|,
\ee
where $\kfb = \Gamma/(\hbar v)$ represents the root-mean-square value of the Fermi momentum at zero chemical potential.

We also define a dimensionless magnetic field strength
\be
B^* \equiv \frac{ e B \ell^2 }{\hbar} = \left( \frac{\ell}{\ell_B} \right)^2,
\ee
where $\ell_B = \sqrt{\hbar/(eB)}$ is the magnetic length.  With these definitions, $\omega_c \tau = (\kfb \ell)^{-1} B^*/P(\rr)$, and we can rewrite the elements of the conductivity tensor in dimensionless form as
\begin{eqnarray}
\frac{\sigma_{xx}(\rr)}{e^2/h} & = &  \frac{\kfb \ell \left| P(\rr) \right|}{1 + \left( \frac{1}{\kfb \ell}\right)^2 \left[\frac{B^*}{P(\rr)} \right]^2}, \nonumber \\
\frac{\sigma_{xy}(\rr)}{e^2/h} & = &  \frac{B^* \textrm{sign}\{P(\rr)\}}{1 + \left( \frac{1}{\kfb \ell}\right)^2 \left[\frac{B^*}{P(\rr)} \right]^2}.
\label{eq:sigmadimensionless}
\end{eqnarray}

Written in this form, the system is characterized by three dimensionless parameters: $B^*$, $\mu/\Gamma$, and $\kfb \ell$.  For a given choice of these parameters, and for a given realization of the random potential $P(\rr)$, one can solve for the current density $\jj(\rr)$ through the system by solving the continuity equation
\be
\del \cdot \jj = 0,
\label{eq:Laplace}
\ee
where $\jj = -\sh \del V(\rr)$, and $V(\rr)$ is the local deviation of the electrochemical potential away from equilibrium.  Below we present results based on a finite element solution to Eq.\ (\ref{eq:Laplace}).  For definiteness, we choose our simulated system to be a Hall bar with length $L = 8\xi$ and width $w = 4 \xi$, where $\xi$ is the correlation length of the potential (equivalent to $r_s$ in the main text).  Applying a unit voltage across the long end of the bar gives the boundary conditions $V(x = 0) = 1$, $V(x = L) = 0$, and $j_y(y = 0) = j_y(y = w) = 0$.  The finite-element mesh size was $\xi/20$.

The random potential was taken to be a Fourier series with random coefficients whose magnitude decreases at high wave vector.  Specifically, we write
\be
P(x,y) = \textrm{Re} \left\{ \sum_{m,n = -\infty}^{\infty} c_{m,n} e^{i (k_m x + k_n y) } \right\} + \frac{\mu}{\Gamma},
\ee
where $c_{m,n}$ are random coefficients and
\be
k_m = \frac{\pi m}{L},  \hspace{5mm} k_n = \frac{\pi n}{w}.
\ee
The coefficients $c_{m,n}$ are chosen to have a random phase in the complex plane and a random magnitude bounded by
\be
|c_{m,n}|^2 \leq e^{-(k_m^2 + k_n^2) \xi^2/2}.
\ee
The normalization of the potential is such that $c_{0,0} = 0$ and $\frac{1}{2} \sum_{m,n} |c_{m,n}|^2 = 1$.

Figure \ref{fig:current_examples} shows a typical numerical solution for the potential $V(x,y)$ and the current $j_x$ in the $x$ direction at $\mu = 0$.  Figures \ref{fig:current_examples}(a)-(b) show the potential and current at zero field, $B^* = 0$, while Figs.\ \ref{fig:current_examples}(c)-(d) show the same system at large field $B^* = 2$.  One can see from these images the strong effect of ``pinch points'' in the random potential, where the current is concentrated at narrow constrictions between puddles.  This focusing of the current becomes more exaggerated at large magnetic field, and correspondingly the electrochemical potential $V(\rr)$ drops abruptly at pinch points and becomes relatively constant far from the boundaries between n- and p-type regions.

For a given realization of the random potential, the longitudinal resistivity can be defined from the simulation as $\rho_{xx} = \Delta V/(L \langle j_x \rangle)$, where $\langle j_x \rangle$ is the area-averaged current density in the $z$ direction and $\Delta V \equiv 1$ is the voltage drop across the system.  One can estimate the dependence of $\rho_{xx}$ on magnetic field $B^*$ by averaging the simulated resistivity $\rho_{xx}$ over many realizations of the random potential for a given value of $B^*$.  Our numerical results, including those shown in Fig.\ 8 of the main text, are averaged over 100 such realizations.  One can also define the Hall resistivity $\rho_{xy}$ by numerically extracting the transverse voltage $V_H$ across the Hall bar at the midpoint, $x = 4\xi$, for a given realization.  The Hall resistivity is defined by $\rho_{xy} = V_H/(w \langle j_x \rangle)$.

In Fig.\ \ref{fig:rhoxx-xy} we show the values of $\rho_{xx}$ and $\rho_{xy}$ as produced by our simulation for a range of magnetic fields $B^*$ and $\mu/\Gamma$.  As in the experimental data (see Fig.\ \ref{Rxy}), $\rho_{xy}$ is much smaller than $\rho_{xy}$ throughout the range of interest.

In order to fit the experimental data at $\mu = 0$ to the numerical simulation, we calculate numerically the resistivity $\rho_{xx}/(h/e^2)$ as a function of $B^*$, using discrete points $B^* = 0, 0.1, 0.2, ..., 2$, for $\kfb \ell = 5, 6$, and $7$.  Linear interpolation allows us to estimate the value of $\rho_{xx}$ for generic values of $\kfb \ell$ between $5$ and $7$ and $B^*$ between $0$ and $2$.  The resulting curves can be translated into real units by inserting the corresponding values of $\kfb$ and $\ell$, which allows us to fit the data shown in Fig.\ 7(b) of the main text.  The result of this fitting is shown in Fig.\ 8(b) of the main text, and corresponds to $\kfb \ell = 6.5$, and $\kfb = 0.88$\,nm$^{-1}$.  This value of $\kfb$ implies a typical electron/hole density $2 \pi \kfb^2 = 1.2 \times 10^{13}$\,cm$^{-2}$, which is consistent in order of magnitude with both our theoretical estimates and our measurements of the Hall constant.

We extracted the dependence of the linear magnetoresistance slope $d[\rho_{xx}/(h/e^2)]/dB^*$ on the chemical potential by numerically evaluating the curve $\rho_{xx}(B^*)$ for different values of $\mu/\Gamma$, and then fitting each curve to a line over the range $0.2 < B^* < 1$.
%
\end{document}